\documentclass[sn-mathphys]{sn-jnl}

\jyear{2022}%

\theoremstyle{thmstyleone}%
\theoremstyle{thmstyletwo}%

\theoremstyle{thmstylethree}%

\begin{document}

\title[43-GHz bandwidth real-time amplitude measurement of 5-dB squeezed]{43-GHz bandwidth real-time amplitude measurement of 5-dB squeezed light using modularized optical parametric amplifier with 5G technology}

\author*[1]{\fnm{Asuka} \sur{Inoue}}\email{asuka.inoue.hy@hco.ntt.co.jp}
\author[1]{\fnm{Takahiro} \sur{Kashiwazaki}}\email{takahiro.kashiwazaki.dy@hco.ntt.co.jp}
\author[2]{\fnm{Taichi} \sur{Yamashima}}\email{yamashima@alice.t.u-tokyo.ac.jp}
\author[2]{\fnm{Naoto} \sur{Takanashi}}\email{takanashi@alice.t.u-tokyo.ac.jp}
\author[1]{\fnm{Takushi} \sur{Kazama}}\email{takushi.kazama.me@hco.ntt.co.jp}
\author[1]{\fnm{Koji} \sur{Enbutsu}}\email{koji.enbutsu.cm@hco.ntt.co.jp}
\author[1]{\fnm{Kei} \sur{Watanabe}}\email{kei.watanabe.hg@hco.ntt.co.jp}
\author[1]{\fnm{Takeshi} \sur{Umeki}}\email{takeshi.umeki.zv@hco.ntt.co.jp}
\author[2,3]{\fnm{Mamoru} \sur{Endo}}\email{endo@ap.t.u-tokyo.ac.jp}
\author*[2,3]{\fnm{Akira} \sur{Furusawa}}\email{akiraf@ap.t.u-tokyo.ac.jp}

\affil[1]{\orgdiv{NTT Device Technology Labs}, \orgname{NTT Corporation}, \orgaddress{\street{3-1, Morinosato Wakamiya}, \city{Atsugi}, \postcode{243-0198}, \state{Kanagawa}, \country{Japan}}}

\affil[2]{\orgdiv{Department of Applied Physics, School of Engineering}, \orgname{The University of Tokyo}, \orgaddress{\street{7-3-1, Hongo}, \city{Bunkyo}, \postcode{113-8656}, \state{Tokyo}, \country{Japan}}}

\affil[3]{\orgdiv{Optical Quantum Computing Research Team}, \orgname{RIKEN Center for Quantum Computing}, \orgaddress{\street{2-1, Hirosawa}, \city{Wako}, \postcode{351-0198}, \state{Saitama}, \country{Japan}}}

\abstract{Continuous-variable optical quantum information processing (CVOQIP), where quantum information is encoded in a traveling wave of light called a flying qubit, is a candidate for a practical quantum computer with high clock frequencies. Homodyne detectors for quadrature-phase amplitude measurements have been the major factor limiting the clock frequency. Here, we developed a real-time amplitude measurement method using a modular optical parametric amplifier (OPA) and a broadband balanced photodiode that is commercially used for coherent wavelength-division multiplexing  telecommunication of the fifth-generation mobile communication systems (5G). 
The OPA amplifies one quadrature-phase component of the quantum-level signal to a loss-tolerant macroscopic level, and acts as a ``magic wand,'' which suppresses the loss after the OPA  from 92.4\% to only 0.4\%.
When the method was applied to a broadband squeezed vacuum with a center wavelength of 1545.32 nm, we observed 5.2 $\pm$ 0.5 dB of squeezing from DC to 43 GHz without any loss correction. The marriage of CVOQIP and 5G technology arranged by the modular OPA will lead to a paradigm shift from the conventional method of using stationary qubits, where the information is encoded in a standing wave system, to a method using flying qubits for ultra-fast practical quantum computation. This means that quantum computer research will move from the stage of developing machines that execute only specific quantum algorithms to a stage of developing machines that can outperform classical computers in running any algorithm.}

\maketitle

\section{Introduction}\label{sec1}
\begin{figure}[h]
\centering
\includegraphics[width=12cm]{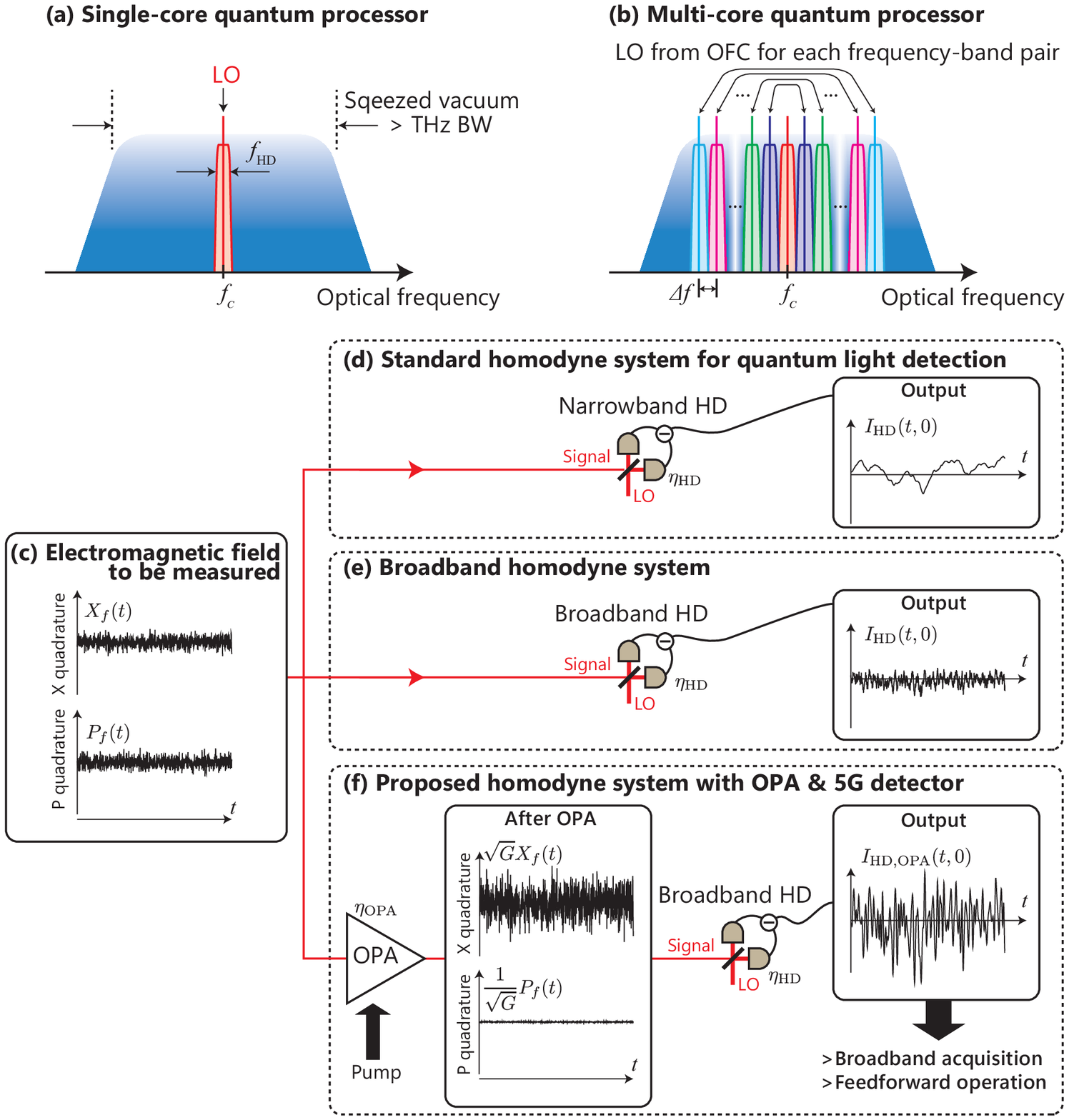}
\caption{(a) Bandwidth (BW) relationship between squeezed vacuum and homodyne detector. $f_\textrm{c}$, center frequency of the squeezed vacuum;$f_\textrm{HD}$, cut-off frequency of the homodyne detector; LO, local oscillator. (b) Concept of a multi-core quantum processor. $\Delta f$, frequency spacing. 
 (c) Quadrature-phase amplitudes $X_f(t), P_f(t)$ of the electromagnetic filed of light to be measured with the temporal mode function $f(t)$. (d-f) Three kinds of homodyne apparatus measuring the $X$ component ($\theta = 0$). Each $I_\textrm{HD}(t,0)$ shows the output signal. The output signals are affected by the detection efficiencies and the bandwidths. (d) Standard homodyne system for quantum light detection. The photodiodes used in the homodyne detector have almost unity detection efficiencies, but the system bandwidth $f_\textrm{HD}$ is typically less than a GHz. (e) Homodyne system with broadband balanced detector, which is often used in 5G applications. The bandwidth $f_\textrm{HD}$ is beyond a GHz but the system efficiency is not high (typically $\sim$50\%).  (f) Proposed homodyne system with an OPA and a balanced detector for 5G applications. The $X$ component of the input light field is amplified by the OPA excited by the pump light and is detected by the homodyne detector with the 5G balanced detector. }\label{fig.1}
\end{figure}

Research on continuous-variable optical quantum information processing (CVOQIP), in which quantum information is encoded in an optical electromagnetic field, is aimed at realizing a fault-tolerant universal quantum computer \cite{Lloyd1999,OBrien2009,Takeda2019,Fukui2022,Asavanant2022}. This form of processing has a high carrier frequency, which can raise the upper limit of the clock frequency. Moreover, its decoherence is small even at room temperature and atmospheric pressure because the photon energy is high compare to the environment.

In CVOQIP, measurement-based quantum computation (MBQC) is a reasonable way of performing large-scale quantum computations \cite{Menicucci2011a,Ukai2015}, rather than circuit-model computations. MBQC uses a cluster state as a resource state, which are special multipartite entangled states.
The cluster state contains a superposition of all input-output relations in quantum computation, and can be regarded as superposition of quantum lookup tables.
In order to select a desired input-output relation from them, that is, to perform the desired computation, we only need to perform local measurements for each qubit by collapsing the wavefunction. It is also called a one-way quantum computation because the measurements perform a quantum calculation and collapse the wavefunction simultaneously. It was initially proposed for a standing wave system, that is, stationary qubit \cite{Raussendorf2001}. In this case, the same qubit must be kept until the calculation is completed, so the lifetime of the qubit, or its memory time, must be longer than the calculation time.

On the other hand, in the case of CVOQIP, the wave packets from the cluster state resource are treated as flying qubits \cite{Lloyd1999}. Since homodyne detectors immediately measure each wave packet, the qubit is consumed, so there is no need to consider the lifetime \cite{Asavanant2021,Larsen2021}. In other words, it can be called a one-pad quantum computation (OPQC).

The processor of an optical OPQC consists of a time-domain multiplexed (TDM) cluster state and a homodyne measurement that measures the quadrature-phase amplitude of each wave packet. The TDM cluster state is created on the basis of a continuous-wave (CW) squeezed vacuum \cite{Yokoyama2013,Yoshikawa2016,Asavanant2019,Larsen2019}.
In a squeezed vacuum, the noise of the electromagnetic field at a particular quadrature-phase component is below the vacuum. A squeezed vacuum is generated by creating quantum correlations between photons contained within a given period often through nonlinear optical processes such as parametric down conversion or four-wave mixing \cite{Andersen2016}.
When two squeezed vacua are interfered with a beamsplitter, quantum entanglement is created between the two spatial modes at the output. The entanglement is time-delayed by an optical delay line and further passed through an asymmetric interferometer to extend the entanglement between time-different wave packets \cite{Menicucci2011a,Ukai2015}.

The broader the squeezing bandwidth is, the shorter the quantum correlation time, defined by the inverse of the bandwidth of the squeezed vacuum, and the more wave packets can be ``packed'' per unit time.  In other words, the upper limit of the clock frequency is the bandwidth of the squeezed vacuum.  It has been reported that low-loss periodically poled lithium niobate (PPLN) waveguides can generate 6-dB of squeezing  in a 6-THz bandwidth with a center wavelength of 1545 nm \cite{Kashiwazaki2021} and 4-dB squeezing in a 25-THz bandwidth with a center wavelength of 2 $\mu$m  \cite{Nehra2022}. That is, a THz clock frequency can be realized in principle.

However, the bandwidth of the homodyne detectors $f_\textrm{HD}$ and other devices is currently limited to about GHz, so the wide bandwidth of the squeezed vacuum is wasted, as depicted in Fig. \ref{fig.1} (a). To measure quantum light, the detection efficiency of a photodiode must be very high, but a photodiode specifically designed for this purpose will inevitably have a narrow bandwidth. Realistically, the bandwidth is limited to a few hundred MHz \cite{Mehmet2011,Huang2013a,Vahlbruch2016a} or at best a few GHz with state-of-the-art silicon photonics technology \cite{Tasker2020,Bruynsteen2021}.

Recently,  a technique employing an optical parametric amplifier (OPA) has been used to measure a broadband squeezed vacuum \cite{Shaked2018,Takanashi2020,Nehra2022}. OPAs act as phase-sensitive optical amplifiers that can amplify one quadrature-phase component without adding noise, i.e., with a noise figure (NF) of 0 dB in principle \cite{Caves1982,Umeki2011, Kazama2021}. This is a significant difference from phase-insensitive optical amplifiers (e.g., erbium-doped fiber amplifiers, Raman amplifiers, semiconductor optical amplifiers, etc.), which cannot cross the NF = 3 dB barrier \cite{Caves1982}. A quantum-level signal can be converted into a macroscopic level with a low-loss OPA with a sufficiently high gain acting as an optical pre-amplifier for quantum measurements \cite{Manceau2017}. Consequently, optical losses and electrical noise are suppressed, which significantly increase the degrees of freedom of the measuring instrument.
However, the previous studies made only narrow bandwidth measurements or power measurements like spectrum analyzers \cite{Shaked2018,Takanashi2020,Nehra2022}, while OPQC requires real-time broadband quadrature-phase amplitude measurements made by homodyne detectors.

In this paper, an OPA amplified one quadrature-phase component of the light to be measured, then we detected its output with a balanced photodiode and a broadband amplifier.
The OPA acts as an optical pre-amplifier and suppresses the losses after the OPA from 92.4\% to only 0.4\%, yielding an overall homodyne measurement system efficiency of 79\%. 
We measured a broadband squeezed vacuum with a center frequency of 194.0 THz (wavelength of 1545.32 nm) generated by another OPA \cite{Kashiwazaki2021}. We observed 5.2-dB squeezing in the band from near DC to 43 GHz without any loss correction. Thus this method can overcome the loss problem, which is the main concern in the field of silicon-photonics based  quantum information processing.

The critical component of this work is a fiber-coupled, low-loss and high-gain broadband OPA with a PPLN waveguide. Many of the other components are commonly used in digital coherent telecommunications in the fifth-generation mobile communication systems (5G).
We believe that the resulting marriage of CVOQIP and 5G technology will lead directly to an optical quantum computer with a clock frequency exceeding 40 GHz.
The 5G elements used in our demonstration were the balanced photodiodes and an electrical amplifier. In addition, as illustrated in Fig. \ref{fig.1} (b), dividing the bandwidth of the squeezed vacuum using a dense wavelength division multiplexing (DWDM) coupler, as is done in  WDM digital coherent communications, enables multiple independent quantum processors to be realized from a single light source. In this case, optical quantum computer can be realized by applying a high-repetition-rate optical frequency comb to the phase-coherent local oscillator light for each homodyne detector.

\section{Results}\label{sec2}
\subsection{Homodyne detection with OPA}
The electromagnetic field of light with a carrier frequency of $\omega_0$ is represented by using annihilation and creation operators, $\hat{A}(t)$ and $\hat{A}^\dagger(t)$, excluding the oscillation term associated with the carrier frequency, as follows:
\begin{align}
\hat{E}(z, t) &\propto \hat{A}(t)e^{i(kz-\omega_0 t)}+\hat{A}^\dagger(t)e^{-i(kz-\omega_0 t)}\\
&=\sqrt{2}\left(\hat{X}(t)\cos(kz-\omega_0 t) + \hat{P}(t)\sin(kz-\omega_0 t)\right).
\end{align}
Here, these operators satisfy the commutation relation $[ \hat{A}(t), \hat{A}^\dagger(t')]=\delta(t-t')$, where $\delta(t)$ is the Kronecker delta function, and $\hat{X}(t)\equiv\frac{\hat{A}(t)+\hat{A}^\dagger(t)}{\sqrt{2}},\hat{P}(t)\equiv \frac{\hat{A}(t)-\hat{A}^\dagger(t)}{\sqrt{2}i}$ are called the quadrature-phase amplitude operators.
We often consider wavepackets of light defined by a temporal mode function $f(t)$ as $\hat{A}_f(t)\equiv\int dt'f(t'-t)\hat{A}(t')$, and the quadrature-phase amplitude operators of this mode are defined as $\hat{X}_f(t)\equiv\frac{\hat{A}_f(t)+\hat{A}_f^\dagger(t)}{\sqrt{2}},\hat{P}_f(t)\equiv\frac{\hat{A}_f(t)-\hat{A}_f^\dagger(t)}{\sqrt{2}i}$. Figure \ref{fig.1} (c) shows examples of measurement results which correspond to eigenvalues $X_f(t)$ and $P_f(t)$ of  $\hat{X}_f(t)$ and $ \hat{P}_f(t)$, respectively.

When we perform balanced homodyne detection using local oscillator (LO) light with an optical frequency of $\omega_0$ as shown in Fig. \ref{fig.1} (d) and (e), the output can be written as
\begin{align}
\hat{I}_\textrm{HD}(t,\theta)\propto\hat{X}_\textrm{HD}(t)\cos\theta + \hat{P}_\textrm{HD}(t)\sin\theta\\
\hat{X}_\textrm{HD}(t) \equiv \int dt'f_\textrm{HD}(t'-t)\hat{X}_f(t')\\
\hat{P}_\textrm{HD}(t) \equiv \int dt'f_\textrm{HD}(t'-t)\hat{P}_f(t'),
\end{align}
where $\theta$ is the phase between the signal and LO lights, and it is assumed that the LO light is sufficiently strong and the quantum fluctuations are negligible. $f_\textrm{HD}$ is the temporal mode function of the instruments, including the homodyne detector, amplifiers, filters, oscilloscope and so on. The Fourier transform of $f_\textrm{HD}(t)$ is related to the frequency response of the measurement system.
For CVOQIP fields, the time-domain data are used not only to read the result but also to process the signal by using feedforward operations.

For example, a  homodyne detector with high detection efficiency ($\eta_\textrm{HD}\sim 1$) is typically used in the case of a standard homodyne system for measuring the quantum state of light (Fig. \ref{fig.1} (d)). Several research groups have used this kind of system to measure high-level squeezed vacua \cite{Mehmet2011,Huang2013a,Vahlbruch2016a} or non-Gaussian states with strong Wigner negativity, such as Schr\"{o}dinger cats \cite{Neergaard-Nielsen2006,Gerrits2010,Asavanant2017,Takase2022} or photon-number states \cite{Neergaard-Nielsen2007a,Yukawa2013}. However, such a homodyne system is specialized for achieving high-detection efficiency, and the bandwidth is not very wide (typically less than GHz order).

In contrast, broadband balanced detectors lift the bandwidth beyond a GHz or even higher. Recently, GHz homodyne detectors built  with silicon photonics technology \cite{Tasker2020,Bruynsteen2021} have been demonstrated; one of them measured a broadband squeezed vacuum \cite{Tasker2020}. Nevertheless, to increase the bandwidth, it is necessary to use photodiodes with a thin active layer and a small photosensitive area, which is not suitable for achieving high detection efficiency. Although the coupling efficiency can be increased by integration, there are still limitations. Losses and electrical noises directly degrade the signal-to-noise ratio (SNR) of the homodyne system, which is critical for CVOQIP applications (Fig. \ref{fig.1} (e)).

Here, we amplify the $X$ component by using an OPA and detect the output with a homodyne detector, as depicted in Fig. \ref{fig.1} (f). The output signal is represented as
\begin{align}
\hat{I}_\textrm{HD,OPA}(t, \theta) \propto \sqrt{G}\hat{X}_\textrm{HD}(t)\cos\theta+\frac{1}{\sqrt{G}}\hat{P}_\textrm{HD}(t)\sin\theta,
\end{align}
where $G$ is the gain of the OPA. Instead of losing information in the $P$ component, the $X$ component is amplified, as shown in Fig. \ref{fig.1} (f). By setting $\theta = 0$, the $X$ component of the quadrature-phase amplitude can be measured. The ideal OPA acts as a phase-sensitive optical amplifier and does not add noise when amplifying the $X$ component, thus suppressing the loss of the broadband homodyne detector.

Let the efficiencies of the homodyne detector and OPA be $\eta_\textrm{HD}$ and $\eta_\textrm{OPA}$, respectively. The effective efficiency of the system can be written as
\begin{align}
\eta_\textrm{eff} = \frac{\eta_\textrm{OPA}\eta_\textrm{HD}}{\eta_\textrm{HD}+\frac{1-\eta_\textrm{HD}}{G}}.\label{eq.eta_eff}
\end{align}
From this expression, if the loss of the OPA is small compared with the loss of the homodyne detector and the gain of the OPA is sufficiently high, $G \gg 1$, the OPA can suppress for the loss of the homodyne detector (see Methods).
Optical losses, the detection efficiency of a broadband balanced photodiode, and electrical noise in the broadband measurement equipment would easily increase the total loss of the homodyne system to more than 90\%, corresponds to $\eta_\textrm{HD} = 10\%$.  Consider, for example, a case where the loss is 90\% ($\eta_\textrm{HD} = 10\%$): if the OPA has a gain $G$ of 35 dB and an efficiency $\eta_\textrm{OPA}$ of 79\% (corresponds to 1-dB NF), the loss after the OPA can be improved from 90\% to only 0.3\%. Moreover, further improvements in efficiency could be achieved in the future if a low-loss and high-gain OPA could be realized.

\subsection{Experimental apparatus}
\begin{figure}[h]
\centering
\includegraphics[width=12cm]{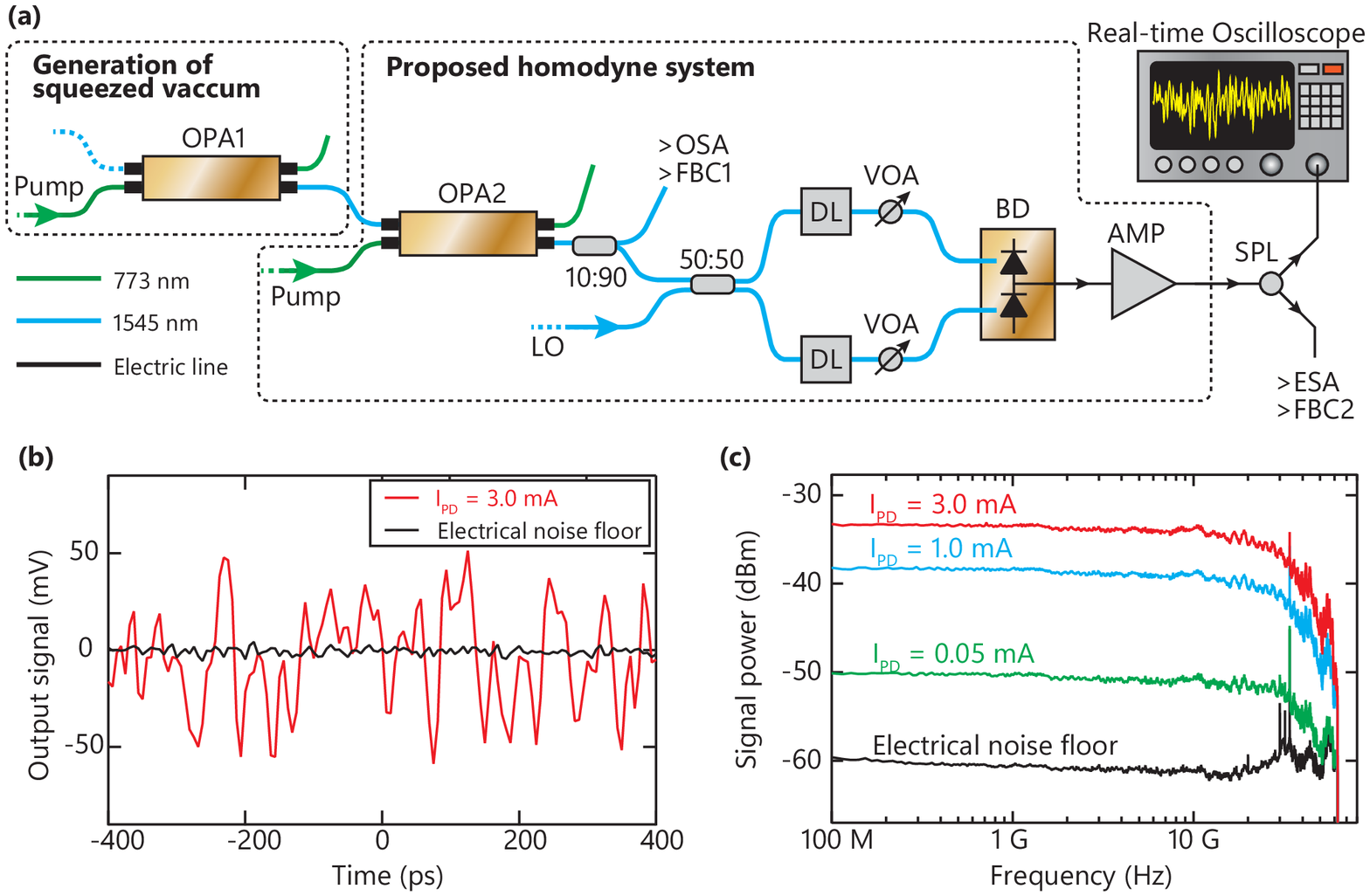}
\caption{(a) Experimental apparatus of the real-time quadrature phase amplitude measurement using the proposed homodyne system. OPA1, optical parametric amplifier used for generating squeezed vacuum; OPA2, optical parametric amplifier for amplifying the$X$ component of the light to be measured; 10:90 and 50:50, fiber-based beamsplitters with coupling ratios of 90\% and 50\%, respectively; LO, local oscillator light (1545 nm); Pump, pump light for the OPAs (773 nm); VOAs, variable optical attenuators; DLs, optical delay lines; BD, balanced photodiode with a bandwidth of 43 GHz; AMP, broadband amplifier; SPL, broadband power splitter; OSA, optical spectrum analyzer; ESA, electrical spectrum analyzer, FBC1, feedback circuit for locking the phase among the squeezed vacuum and the pump light at OPA2; FBC2, feedback circuit for locking the phase at the amplified signal and LO. (b) Output signals measured by a real-time oscilloscope. The black trace shows the electrical noise floor when all lights are blocked. The red trace shows the shot noise of the LO when the photocurrent of a single photodiode is 3.0 mA. (c) Frequency responses of several photocurrents calculated by fast Fourier transformation of the oscilloscope data.}\label{fig.2}
\end{figure}

Figure \ref{fig.2} (a) shows the experimental apparatus of our homodyne system. The fundamental light is from a continuous-wave fiber laser (NKT photonics, X15) at 1545.32 nm (not shown).
The light to be measured is a broadband squeezed vacuum generated by a fiber pig-tailed OPA (OPA1 in the figure) \cite{Kashiwazaki2021} . The output fiber of OPA1 is connected to the input fiber of OPA2 (used as a phase-sensitive optical amplifier), which is excited by pump light with a wavelength of 773 nm. 10\% of the OPA2 output is used for monitoring the output spectrum by an optical spectrum analyzer (OSA, Yokogawa AQ6370D), and locking the phase between the squeezed vacuum and the sub harmonic of the pump light by a feedback circuit (FBC1) \cite{Takanashi2020} (see the supplementary information).
The remaining 90\% is sent to a 50:50 fiber beamsplitter, where it interferes with the local oscillator (LO) light and is received by a balanced photodiode (BD, $\textrm{u}^2\textrm{t}$, BPDV21x0R). Delay lines (DLs) and variable optical attenuators (VOAs) are inserted in each path to adjust the optical path length and power balance. The two photodiodes are reverse biased by biasing circuits, and each photocurrent $I_\textrm{PD}$ is monitored (not shown). The output of the balanced photodiode is amplified by a broadband amplifier (AMP, SHF, S807 B, 55-GHz bandwidth). The output is split by a broadband power splitter (SPL, Anritsu, W241A); then, one part is measured by a real-time oscilloscope (Keysight, DSO-Z 634A, 63-GHz bandwidth). The other part is monitored by an electrical spectrum analyzer (ESA, Keysight, EXA N9010B). The ESA is also used to generate an error signal to lock the phase between the LO light and the output of OPA2 by a feedback circuit (FBC 2) (see the supplementary information for details).

\subsection{Characterization of the homodyne system}
In the following experiments, the pump power of OPA2 was set to 1.2 W. The parametric gain $G$ and NF of OPA2 were 35 dB and 1 dB, respectively (see the supplementary information and the reference \cite{Kazama2021}). The measured loss after the OPA2 was 92.4\%, then from Eq.\ref{eq.eta_eff} the effective efficiency of the total homodyne system is 79\%.
Figure \ref{fig.2} (b) shows the raw data of the homodyne detector. The black trace shows the electrical noise floor when all the light is blocked. The red trace shows the shot noise with the LO when the photocurrent of each photodiode is set to 3.0 mA. We acquired the output signal for 78.2 ns. Figure \ref{fig.2} (c) shows fast Fourier transform (FFT) traces of the acquired data with different LO powers. These traces were calculated by averaging the results of 8192 measurements taken over a 0.8-ms period and all the frequency response traces are averaged for 8192 traces. The traces show a flat frequency response up to 10 GHz. The 3-dB bandwidth of our system is 43 GHz of the BD, and even at that frequency, 20 dB of SNR is obtained when the photocurrent is 3.0 mA (red trace). All the traces are fall off beyond the frequency of 63 GHz, which is the the edge of the oscilloscope's bandwidth. Note that OPA2 has a THz bandwidth, which means a much wider bandwidth can be obtained if faster electronics can be used.

\subsection{Measurement of broadband squeezed light}
\begin{figure}[h]
\centering
\includegraphics[width=12cm]{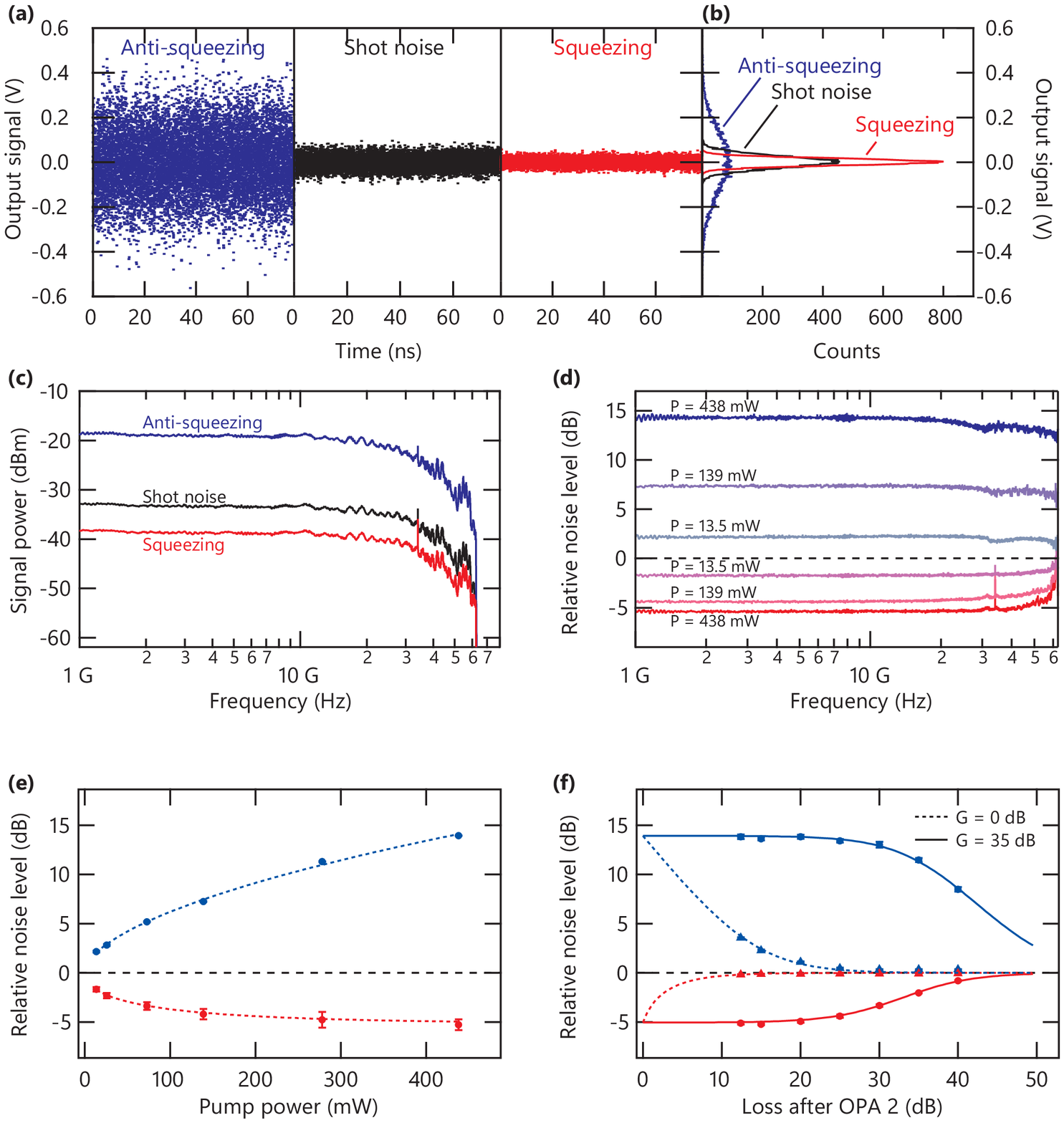}
\caption{Measurement results of the broadband squeezed vacuum. (a) Raw data acquired by the oscilloscope. Each point represents the quadrature phase component. (left) Anti-squeezing. (center) Shot noise. (right) Squeezing. (b) Histograms calculated from (a). (d) Frequency domain data calculated by FFT from the time-domain data. Each trace is averaged for 8192 frames of 12512 points of data. The pump power for OPA1 is 438 mW. (d) Relative squeezing and anti-squeezing levels in the frequency domain for several pump powers. (e)Pump-power dependence of the squeezing (red circles) and anti-squeezing (blue circles) levels. The dashed lines show fitting results. (f) Calculated (lines) and measured (markers) squeezing levels with additional loss after OPA2. The gain of OPA2 was set to 0 dB (dashed lines) and 35 dB (solid lines). }\label{fig.3}
\end{figure}

Figure \ref{fig.3} (a) shows the raw data corresponding to quadrature-phase amplitudes of the squeezed light of the anti-squeezing and squeezing components (left and right, respectively). The center shows the shot noise when the pump for OPA1 is set to 0 mW. Figure \ref{fig.3} (b) is a histogram based on the measurement results of Fig. \ref{fig.3} (a).
Figure \ref{fig.3} (a) and (b) clearly show quadrature squeezing of the light to be measured.

The FFTs of these data are shown in Fig. \ref{fig.3} (c), and the normalized squeezing level calculated for several pump powers of OPA1 is shown in Fig. \ref{fig.3} (d).
From these figures, it can be seen that the homodyne system successfully measured a 5.2-dB broadband squeezed vacuum from DC  to 43 GHz. Note that the peaks at 34 GHz are measurement artifacts caused by the oscilloscope.

Figure \ref{fig.3} (e) shows the squeezing and anti-squeezing levels as the function of pump power for OPA1. The occupied circles are calculated from the variance of time-domain data like in Fig. \ref{fig.3} (a), and the error bars are estimated from 8192 data sets. For a pump power of 438 mW, a squeezing level of $5.2\pm0.5$ dB and an anti-squeezing level of $13.9\pm0.1$ dB are obtained. The dashed line is the fitting to the function:
\begin{align}
R_{\pm}(P) =  L + (1-L)\exp\left(\pm2\sqrt{aP}\right)
\end{align}
where $\pm$ represents anti-squeezing (+, blue) or squeezing (-, red), $L$ represents the total effective loss of the system, including of OPA1 and the homodyne system, and $a$ is a second-harmonic-generation coefficient in the unit of $\textrm{W}^{-1}$.
The fitting result shows the system loss of 29\%, including losses of OPA1, OPA2, optical propagation, detector, and the electrical noise of the instruments.
This value limits the squeezing level that can be measured with this apparatus to $5.2$ dB.

In order to confirm the performance of the suppressing  losses after OPA2, we inserted a VOA (not shown in Fig. \ref{fig.2} (a)) after OPA2 to worsen $\eta_\textrm{HD}$ and measured the squeezing level as a function of the loss as shown in Fig. \ref{fig.3} (f). Note that the measurements in Fig. \ref{fig.3} (f) are not broadband measurements by the oscilloscope but rather narrowband measurements made with ESA at a sideband frequency of 100 MHz and we did not lock the LO phase (see supplementary information).
The dashed and solid lines are calculated traces when $G=0$ and $35$ dB. The markers with error bars represent the measured values. 
It is clear from this figure that OPA2 with $G=35$ dB effectively suppresses the loss after OPA2 from 92.4\% to 0.4\%, and the system can measure the original squeezing level.

\section{Discussion}\label{sec3}
The most significant advantage of optical quantum computers is that the upper limit of the clock frequency is by far the highest compared with other methods. The bandwidth of the squeezed light used to generate the cluster states is above THz, meaning that a THz-clocked quantum computer can be realized in principle. However, the bandwidth on the measurement instruments, including the homodyne detectors, has not kept pace, and a large part of the bandwidth of the squeezed light is wasted. We have experimentally demonstrated a high-detection efficiency (79\%) and wide-bandwidth (43 GHz) homodyne system using an OPA module and commercially available 5G components. The OPA drastically suppress the losses after OPA2 from 92.4\% to only 0.4\%.  With this system, we performed real-time measurements of the quadrature-phase amplitude of broadband squeezed vacuum, obtaining a squeezing level of 5.2$\pm$0.5 dB without loss correction. This value exceeds one of the thresholds (4.5 dB) for cluster state generation \cite{Asavanant2019}, indicating that this method can be applied to quantum calculations. In fact, except for the handling of non-Gaussian quantum states \cite{Gottesman2001,Fukui2022}, the necessary elements for optical quantum computation are LO phase switching and feedforward systems, which are parts of classical information processing. Even at present, there are electro-optical modulators and electronics above 40 GHz in bandwidth as 5G technology \cite{Wang2018a,Pittala2022}, and when ``Beyond 5G'' or 6G technologies \cite{Dang2020} appear in the future, the speed of classical information processing will be further increased. This means that the broadband homodyne detector developed in this study will make it feasible for the clock frequency of optical quantum computers to exceed 40 GHz.

We conclude this paper by proposing a multi-core optical quantum processor by effectively using the full-bandwidth of squeezed vacuum with our homodyne system.

Quantum entanglement between two frequency sidebands is the essential feature of a squeezed vacuum. In this study, we measured the quantum correlation  in the $\pm$43 GHz bandwidth around a frequency $f_\textrm{c}$ of 194.0 THz, as shown in Fig. \ref{fig.1} (a). Therefore, by dividing the optical spectrum into frequency-band pairs with DWDM couplers (e.g., with a frequency spacing $\Delta f=100$ GHz) and applying our homodyne system for each as illustrated in Fig. \ref{fig.1} (b), a multi-core optical quantum processor can be realized, wherein the operating clock of each processor exceeds 40 GHz. As well, optical frequency comb technology can be used to prepare phase-coherent LO beams for each homodyne system \cite{Yang2021,Corcoran2020,Chang2022}. This research shows that CVOQIP is highly compatible with mature 5G technology. 

Increasing the clock frequency and core multiplexing technologies have dramatically enhanced the processing speed of classical computers, which created the current information society. The method proposed in this paper shows that high-clock frequency and multi cores are also possible in optical quantum computers. Ultra-fast multi-core optical quantum processors will catapult quantum computer research into a new era, where versatile quantum computers surpass classical computers in any algorithm, not just in specific quantum algorithms.

\section{Methods}
\subsection{OPAs}
We used two kinds of OPA. Both are 45-mm-long, ZnO-doped PPLN waveguides, but their fabrication processes and waveguide structures are different. This is because their different applications require modules with different characteristics. 
OPA1, which is used for generating the broadband squeezed vacuum, is fabricated by mechanically polishing and the thickness and the width of the waveguide are about 8 $\mu$m and 8.6 $\mu$m. Thanks to the smooth surface and relatively large waveguide core, the waveguide loss is only 7\% and even the total loss of the fiber pig-tailed module is only 21\%, which is an appropriate level for generating squeezed vacuum. In contrast, the parametric gain of this module is not very high ($\sim$15 dB when the pump power of 438 mW).

In contrast, OPA2 is fabricated by a chemical etching process, and its thickness and width are about 5 $\mu$m and 5.6 $\mu$m. Because the waveguide is relatively smaller than the one of OPA1, the loss is slightly higher, but a parametric gain of 35 dB can be obtained for a pump power of 1.2 W. 
See the references for more details (OPA1 \cite{Kashiwazaki2021}, OPA2 \cite{Kashiwazaki2020,Kazama2021}).

\subsection{Effective efficiency}
Equation \ref{eq.eta_eff} can be re-written as follows with efficiencies when $\theta = 0$:
\begin{align}
\hat{I}_\textrm{HD,OPA}(t, 0) \propto &\sqrt{\eta_\textrm{HD}\eta_\textrm{OPA}G}\hat{X}_\textrm{HD}(t)\notag\\ 
&+\sqrt{\eta_\textrm{HD}(1-\eta_\textrm{OPA})G}\hat{X}_\textrm{vac,OPA}(t)+\sqrt{1-\eta_\textrm{HD}}\hat{X}_\textrm{vac,HD}(t),
\end{align}
where $\hat{X}_\textrm{vac,OPA}(t)$ and $\hat{X}_\textrm{vac,HD}(t)$ are the quadrature-phase amplitude of the vacuum due to the losses.
From this equation, the effective efficiency $\eta_\textrm{eff}$ is calculated as 
\begin{align}
\eta_\textrm{eff}=&\frac{\eta_\textrm{HD}\eta_\textrm{OPA}G}{\eta_\textrm{HD}\eta_\textrm{OPA}G + \eta_\textrm{HD}(1-\eta_\textrm{OPA})G+(1-\eta_\textrm{HD})}\\
&=\frac{\eta_\textrm{OPA}\eta_\textrm{HD}}{\eta_\textrm{HD}+\frac{1-\eta_\textrm{HD}}{G}}\\
&\xrightarrow{G\to\infty}\eta_\textrm{OPA}.
\end{align}

\section{Data availability}
Data are available from the authors upon request.

\section{Acknowledgments}
The authors acknowledge supports from the UTokyo Foundation and donations from Nichia Corporation of Japan. T.Y. acknowledges support from the Advanced Leading Graduate Course for Photon Science (ALPS). M.E. acknowledges support from the Research Foundation for Opto-Science and Technology. This work was partly supported by the Japan Science and Technology Agency (JPMJMS2064) and the Japan Society for the Promotion of Science KAKENHI (18H05207,  20K15187). The authors acknowledge Dr. Hiroshi Yamazaki for supporting the setup of the broadband detection system, and Mr. Takeru Ebihara for fruitful discussion. 

\section{Author contributions}
A.I. N.T., and T.U. conceived and planned the project.
A.I., T. Kashiwazaki, T.Y. and T.U. designed and constructed the experimental apparatus and acquired the data.
T. Kashiwazaki, T. Kazama, K.E., K.W., and T.U. developed the low-noise and high-gain OPA module.
A.I., T. Kashiwazaki, and M.E. analyzed the data and wrote the manuscript with assistance from all other co-authors.
A.F. supervised the project.
\section{Competing interests}
The authors declare no competing financial interests.

\bibliography{40GHz}

\end{document}